\documentclass[12pt]{article}
\usepackage{amsfonts,amssymb,amsmath,eucal,amsthm,color}
\usepackage[dvips]{graphics}

\begin{document}
\textheight=8.5truein
\textwidth=6.1truein
\overfullrule=0pt
\parskip=2pt
\parindent=12pt
\headheight=0in
\headsep=0in
\topmargin=0.5in
\oddsidemargin=0in
\newsavebox{\ns}
\newsavebox{\dbrane}
\newsavebox{\dbshort}
\renewcommand{\arraystretch}{1.2}
\newcommand{\sect}[1]{\setcounter{equation}{0}\section{#1}}
\renewcommand{\theequation}{\arabic{section}.\arabic{equation}}
\def\appendix{{\newpage\section*{Appendix}}\let\appendix\section%
        {\setcounter{section}{0}
        \gdef\thesection{\Alph{section}}}\section}
\def\theequation{\thesection.\arabic{equation}}
\def\be{\begin{equation}}
\def\ee{\end{equation}}
\def\ba{\begin{eqnarray}}
\def\ea{\end{eqnarray}}
\newcommand{\nn}{\nonumber}
\newcommand\para{\paragraph{}}
\newcommand{\ft}[2]{{\textstyle\frac{#1}{#2}}}
\newcommand{\eqn}[1]{(\ref{#1})}
\newcommand\balpha{\mbox{\boldmath $\alpha$}}
\newcommand\bbeta{\mbox{\boldmath $\beta$}}
\newcommand\bgamma{\mbox{\boldmath $\gamma$}}
\newcommand\bomega{\mbox{\boldmath $\omega$}}
\newcommand\blambda{\mbox{\boldmath $\lambda$}}
\newcommand\bmu{\mbox{\boldmath $\mu$}}
\newcommand\bphi{\mbox{\boldmath $\phi$}}
\newcommand\bzeta{\mbox{\boldmath $\zeta$}}
\newcommand\bsigma{\mbox{\boldmath $\sigma$}}
\newcommand\bepsilon{\mbox{\boldmath $\epsilon$}}
\def\Dslash{\,\,{\raise.15ex\hbox{/}\mkern-12mu D}}
\def\Dbarslash{\,\,{\raise.15ex\hbox{/}\mkern-12mu {\bar D}}}
\def\delslash{\,\,{\raise.15ex\hbox{/}\mkern-9mu \partial}}
\def\delbarslash{\,\,{\raise.15ex\hbox{/}\mkern-9mu {\bar\partial}}}
\def\pslash{\,\,{\raise.15ex\hbox{/}\mkern-9mu p}}
\def\calDslash{\,\,{\raise.15ex\hbox{/}\mkern-12mu {\cal D}}}
\newcommand\Bprime{B${}^\prime$}
\newcommand\sign{{\rm sign}}
\newcommand\mf{\mathcal{F}}
\newcommand\ma{\mathcal{A}}
\newcommand\mq{\mathcal{Q}}
\newcommand\mr{\mathbb{R}}
\newcommand\mcp{\bf \mathbb{C}P}
\newcommand\mhp{\bf \mathbb{H}P}
\newcommand\Sfour{\Sigma^{-}{\bf S}^4}
\newcommand\Sfive{\mathbb{R}^3 \times {\bf S}^5}
\newcommand\mh{\mathbb{H}}
\newcommand\ind{\mathrm{ind}}
\newcommand\Index{\mathrm{Index}}
\newcommand\myOverwrite[2]{\makebox[0cm][l]{#1}#2\ }
\newcommand\D{\myOverwrite{D}{\slash}}
\newcommand{\Mb}{\overline{\mathcal{M}}}
\newcommand{\C}{\mathbf{C}}
\newcommand{\R}{\mathbf{R}}
\newcommand{\Z}{\mathbf{Z}}
\newcommand{\T}{\mathbf{T}}
\renewcommand{\P}{\mathbf{P}}
\newcommand{\cO}{\mathcal{O}}
\begin{titlepage}
\begin{center}
\today
{\small\hfill hep-th/0309208}\\
{\small\hfill HUTP-03/A061}\\
\vskip 1.5 cm
{\large \bf Quantum Calabi-Yau and Classical Crystals} \\
\vskip 1.2 cm
{Andrei Okounkov$^1$, Nikolai Reshetikhin$^2$, and Cumrun Vafa$^3$}\\
\vskip .4cm
\vskip 0.4cm
$^1${\sl Department of Mathematics, \\
Princeton University, Princeton, NJ 08544, U.S.A. \\
{\tt okounkov@math.princeton.edu}\\}
\vskip 0.5cm
$^2${\sl Department of Mathematics,\\
 University of California at Berkeley,
Evans Hall, \# 3840, Berkeley, CA 94720-3840, U.S.A. \\
 {\tt reshetik@math.berkeley.edu}\\}
\vskip 0.5cm
$^3${\sl Jefferson Physical Laboratory, Harvard University, \\
Cambridge, MA 02138, U.S.A. \\ {\tt vafa@string.harvard.edu}\\}
\end{center}
\vskip 0.2 cm
\begin{abstract}
We propose a new duality involving topological
strings in the limit of large string coupling constant.
  The dual is described
in terms of a classical statistical mechanical model
of crystal melting, where the temperature is inverse
of the string coupling constant.  The crystal is a discretization
of the toric base of the Calabi-Yau with lattice length
$g_s$.  As a strong evidence for this duality
we recover the topological vertex in terms of the statistical
mechanical probability distribution for crystal melting. 
 We also propose a more
general duality involving the dimer problem on periodic lattices
and topological A-model string on arbitrary local toric threefolds. 
The $(p,q)$ 5-brane web, dual to Calabi-Yau,
gets identified with the transition regions of rigid
dimer configurations.
\end{abstract}
\end{titlepage}
\pagestyle{plain}
\setcounter{page}{1}
\newcounter{bean}
\baselineskip16pt
\tableofcontents
\section{Introduction}
Topological strings on Calabi-Yau threefolds
 have been a fascinating class of string
theories, which have led to insights
into dynamics of superstrings and supersymmetric
gauge theories.  They
have also been shown to be equivalent in some cases to non-critical
bosonic strings.  In this paper we ask how the topological
A-model which `counts' holomorphic curves inside the Calabi-Yau
behaves in the limit of large values of the string coupling
constant $g_s\gg1$.  We propose a dual description which
is given in terms of a discrete statistical mechanical model of a
three dimensional real crystal with boundaries, where 
the crystal is located in the toric base of the Calabi-Yau
threefold, with the `atoms' separated
by a distance of $g_s$. Moreover the Temperature $T$ in the statistical
mechanical model corresponds to $1/g_s$.  Heating up the
crystal leads to melting of it.  In the limit of large temperature,
or small $g_s$, the Calabi-Yau geometry emerges from the geometry
of the molten crystal !

In the first part of this paper we focus on the simplest Calabi-Yau, namely ${\bf C}^3$.
Even here there are a lot of non-trivial questions to answer.
In particular the computation of topological string amplitudes when
we put D-branes in this background is non-trivial and leads
to the notion of a topological vertex 
\cite{Aga,iqb} (see also the recent paper
\cite{Dia}).  Moreover
using the topological vertex one can compute an all order amplitude
for topological strings on arbitrary local Calabi-Yau manifolds.  This
is an interesting class to study, as it leads to non-trivial predictions
for instanton corrections  to gauge and gravitational
couplings of a large class of $N=2$ supersymmetric
gauge theories in 4 dimensions via geometric engineering 
\cite{kkv} (for recent progress
in this direction see \cite{nek,iqbal}).   It is also
the same class which is equivalent (in some limits) to non-critical
bosonic string theories.
 In this paper
we will connect the topological vertex to the partition function 
of a melting corner with fixed asymptotic 
boundary conditions.  Furthermore we find an intriguing
link between dimer statistical mechanical models and non-compact
toric Calabi-Yau threefolds.  In particular the dimer problems
in 2 dimensions naturally get related to the study of configurations
of $(p,q)$ 5-brane web, which is dual to non-compact toric Calabi-Yau
threefolds.

The organization of this paper is as follows:  In Section 2
we will motivate and state the conjecture.  In Section 3 
we check aspects of this
conjecture and derive the topological vertex from the 
statistical mechanical model.
In Section 4 we discuss dimer problems
and its relation to topological strings on 
Calabi-Yau.

Our proposal immediately raises many questions, which 
are being presently investigated. Many of them will 
be pointed out in the paper. One of the most
interesting physical questions involves the superstring
intepretation of the discretization of space.  On the
mathematical side, we expect
that our statistical mechanical model 
should have a deep meaning in terms of the 
geometry of target space based on the interpretation of 
its configurations as torus fixed points in the Hilbert
scheme of curves of the target threefold. Also, our 
3d model naturally extends the 
random 2d partition models that arise in $N=2$
supersymmetric gauge theory \cite{nek} and Gromov-Witten 
theory of target curves \cite{OP}.  For some mathematical
aspects of the topological vertex see \cite{Liu,OP2,Dia}.
\section{The Conjecture}
\subsection{Hodge integrals and 3d partitions}
Consider topological A-model strings on a Calabi-Yau threefold.  
For simplicity, 
let us consider the limit when the K\"ahler class of 
Calabi-Yau is rescaled by a factor that goes to infinity.
As explained in \cite{bcov}\ in this limit the genus $g$ amplitude
is given by 
$$
\frac{\chi}2 \int_{\Mb_g} c_{g-1}^3({\mathcal{H}})
$$ 
where $\chi$ is the Euler characteristic of the Calabi-Yau
threefold, $\Mb_g$ is the moduli space of Riemann surfaces, 
${\cal H}$ is the Hodge bundle over $\Mb_g$, and 
$c_{g-1}$ denotes
the $(g-1)$-st Chern class of it.
  For genus 0 and 1, there are also some K\"ahler dependence (involving
volume and the second Chern class of the tangent bundle), which we subtract out
to get a finite answer.  Consider
$$
Z=\exp\left[\frac{\chi}2 \sum_g g_s^{2g-2} \int_{\Mb_g} c_{g-1}^3({\cal H})
\right] \,.
$$
It has been argued physically \cite{gv}\ and derived mathematically 
\cite{fp}\ that 
$$Z=f^{\chi/2}$$
where
$$f=\prod_n{1\over (1-q^n)^n}$$
and
$$q=e^{-g_s}\,.$$
By the classical result of McMahon, the function $f$ is
the generating function for 3d partitions, that is,
$$
f=\sum_\textup{3d partitions} q^{\textup{\# boxes}}\,,
$$
where, by definition, a 3d partition is a 3d generalization
of 2d Young diagrams and is an object of 
kind seen on the left in Figure \ref{f3}. This
fact was pointed out to one of us as a curiosity by
R.~Dijkgraaf  shortly
after \cite{gv}\ appeared.  
\subsection{Melting of a crystal and Calabi-Yau threefold}\label{MCY}
 It is natural to ask whether there is a deeper
reason for this correspondence.
 What could three dimensional partitions 
have to do with A-model topological string on a Calabi-Yau threefold ?
The hint comes from the fact that we are considering the limit of large K\"ahler
class and in this limit the Calabi-Yau looks locally made of $\C^3$'s 
glued together.  It is then natural
to view this torically, as we often do in topological string, in the context of
mirror symmetry, and write the K\"ahler form as
$$\sum_{i=1,2,3} dz_i\wedge d{\overline z}_i\sim \sum_{i=1,2,3} d|z_i|^2 \wedge d\theta_i$$
and $|z_i|^2 $ span the base of a toric fibration of $\C^3$.  
Note that $|z_i|^2=x_i$ parameterize
the positive octant $O^+\subset \R^3$.  
If we assign Euler characteristic ``2'' to each
$\C^3$ patch, the topological string amplitudes on it
get related to the McMahon function.
Then it is natural to think that the octant is related to the three dimensional partitions, in which the boxes are located at 
$\Z^3$ lattice points inside $O^+$. Somehow the points of the Calabi-Yau,
in this case $\C^3$,
become related to integral lattice points on the toric base.  

\begin{figure}[!hbtp]
  \begin{center}
    \scalebox{0.4}{\includegraphics{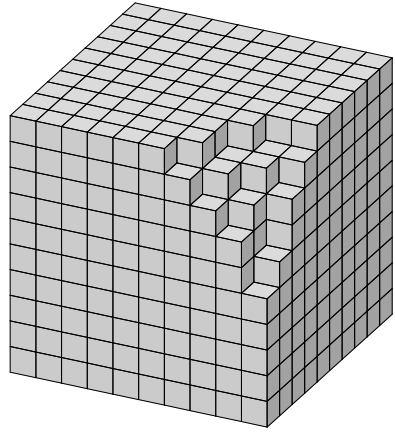}} 
    \caption{A melting crystal corner}
    \label{f1}
  \end{center}
\end{figure}

The picture we propose is the following. We identify the 
highly quantum Calabi-Yau with the frozen crystal, that is, 
the crystal in which all atoms (indexed by lattice points
in $O^+$) are in place.
We view the 
excitations as removing lattice points as in Figure \ref{f1}. 
The rule is that we can remove lattice points only if there
are no pairs of atoms on opposite sides.  This gives the same
rule as 3d partitions.
Note that if one holds the page upside-down, one sees a 3d
partition in Figure \ref{f1}, namely the partition from 
Figure \ref{f3}. 

\begin{figure}[!hbtp]
  \begin{center}
    \scalebox{0.64}{\includegraphics{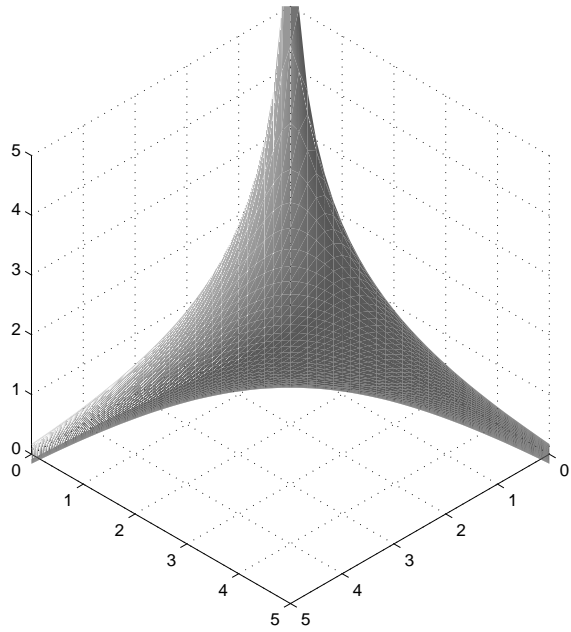}} 
    \caption{The limit shape of a 3d partition}
    \label{f2}
  \end{center}
\end{figure}

Removing each atom contributes the factor $q=e^{-\mu/T}$  to the 
Boltzmann weight of the configuration, where $\mu$ is the chemical
potential (the energy of the removal of an atom) and
$T$ is the temperature. We choose units in which $\mu=1$.   
To connect this model to topological string on $\C^3$ we identify $g_s=1/T$.
In particular, the $g_s\to 0$, that 
is, the $q\to 1$ limit the crystal begins to melt 
away. Rescaled in all direction by a factor of $1/T=g_s$,
the crystal approaches a smooth limit shape which 
has been studied from various viewpoints \cite{CK,OR}. 
It is plotted in Figure \ref{f2}. 

The analytic form of this limit shape is 
encoded in terms of a complex
Riemann surface, which in this case is given by
\begin{equation}
  \label{curveF}
  F(u,v)=e^{-u}+e^{-v}+1=0
\end{equation}
defined as a hypersurface in $\C^2$ with a natural
2-form $du\wedge dv$, where $u,v$ are periodic variables with
period $2\pi i$. In coordinates $e^{-u}$ and $e^{-v}$ this 
is simply a straight line in $\C^2$. Consider the 
following function of the variables $U=Re(u)$ and $V=Re(v)$
\begin{equation}
  \label{Rnk}
  R(U,V) = \frac1{4\pi^2} \iint_{0}^{2\pi} 
\log \, |F(U+ i \theta, V+i \phi)|\, d\theta d\phi\,.
\end{equation}
This function is known as the \emph{Ronkin function} of $F$. 
In terms of the Ronkin function, the limit shape can 
be parameterized as follows: 
$$
(x_1,x_2,x_3)=(U+R,V+R,R)\,,\quad R=R(U,V) \,. 
$$
Note that 
$$
U = x_1 - x_3 \,, \quad V = x_2  - x_3 \,.
$$
The projection of the curved part of the
limit shape onto the $(U,V)$-plane is the
region bounded by the curves 
$$\pm e^{-U} \pm e^{-V}+1=0\,,$$
excluding the case when both signs are positive. 
This region is the \emph{amoeba} of the curve 
\eqref{curveF}, which, by definition, is its
image under the map $(u,v)\mapsto (U,V)$. 
In different coordinates, this is the planar
region actually seen in Figure \ref{f2}. 
\subsection{Mirror symmetry and the limit shape}
Consider topological strings on ${\bf C}^3$.  One
can apply mirror symmetry in this context by dualizing
the three phases of the complex parameters according
to T-duality.  This has been done in \cite{hv}.
One introduces dual variables $Y_i$ which are periodic,
with period $2\pi i$ and are related to the $z_i$ by 
$$|z_i|^2=Re Y_i$$
(this is quantum mechanically modified by addition
of an $i$-independent large positive constant to the right hand side).
The imaginary part of the $Y_i$ is `invisible' to the original
geometry, just as is the phase of the $z_i$ to the $Y_i$
variables.  They are T-dual circles.  But we can compare
the data between ${\bf C}^3$ and the mirror on the base
of the toric variety which is visible to both.  In mirror
symmetry the K\"ahler form of ${\bf C}^3$ 
gets mapped to the holomorphic three form which in this
case is given by
$$\Omega =\prod_{i=1}^3 dY_i  \ {\rm exp}[ W]$$
where 
$$W=\sum e^{-Y_i}.$$
Note that shifting $Y_i\rightarrow Y_i+r$ shifts
$W\rightarrow e^{-r} W$. The analog of rescaling
in the mirror is changing the scale of $W$.  Let us
fix the scale by requiring $W=1$; this will turn out
to be the mirror statement to rescaling by $g_s$ to
get a limit shape.  In fact we will see below
that the limit shape corresponds to the toric
projection of the complex surface
in the mirror given simply by $W(Y_i)=1$.
Let us define
$$u=Y_1-Y_3, \quad v=Y_2-Y_3$$
Then 
$$W=e^{-Y_3} F(u,v)$$
where
$$F(u,v)=e^{-u}+e^{-v}+1$$
we will identify $F(u,v)$ with the Riemann surface of
of the crystal melting problem.  In this context also
according to mirror map the points
in the $u,v$ space get mapped
to the points on the toric base
(i.e $O^+$) satisfying
$$x_1-x_3=Re (u)=U$$
$$x_2-x_3=Re (v)=V$$
We now wish to understand the interpretation
of limit shape from the viewpoint of topological
string.  We propose that the
 boundary of the molten crystal which is a 2-cycle
on the octant should be viewed as a special Lagrangian
cycle of the A-model with one hidden circle in the fiber.
Similarly in the B-model mirror it should be viewed as the
B-model holomorphic surface, which in the case at hand
gets identified with $W=1$ (recall that in the LG models
B-branes can be identified with $W={\rm const.}$ \cite{hiqv}). 
 On this surface we have
$$e^{-Y_3}F(u,v)=1$$
If we take the absolute value of this equation, to
find the projection on to the base we find
$$e^{-Re(Y_3)}|F(u,v)|=1$$
If we take the logarithm of this relation we have
$$-Re(Y_3)+{\rm log}|e^{-u}+e^{-v}+1|=0\rightarrow
x_3={\rm log}|e^{-u}+e^{-v}+1|
$$
However we have a fuzziness in mapping this to the toric base:
$u=Re(u)+i \theta$ and $v=Re(v)+i \phi$ and so a given
value for $Re(u)$ and $Re(v)$ does not give a fixed value
of $x_3$.  That depends in addition on the angles $\theta, \phi$
of the mirror torus which are invisible to the A-model toric base.
It is natural to take the average values as defining the projection
to the base, i.e.
$$x_3={1\over 4\pi^2}\int d \theta d\phi \ {\rm log}|
F(U+i\theta, V+i \phi)|$$
which is exactly the expression for the limit shape.
We thus find some further evidence that the statistical
mechanical problem of crystal melting is rather deeply
related to topological string and mirror symmetry on Calabi-Yau.  Moreover we
can identify the points of the crystal, with the
discretization of points of the base of the toric Calabi-Yau.

To test this conjecture further we will have to first
broaden the dictionary between the two sides.  In particular
we ask what is the interpretation of the topological vertex
for the statistical mechanical problem of crystal melting?
For topological vertex we fix a 2d partition on each of the
three legs of the toric base.  There is only one natural
interpretation of what this could mean in the crystal melting
problem: This could be the partition function of the melting crystal
with three fixed asymptotic
boundary shapes for the molten crystal,
dictated by the corresponding partition.  We will show
this is indeed the case in the next section. 
\section{Melting corner 
and the topological vertex}
\subsection{Transfer matrix approach}
The grand canonical ensemble of 3d partitions weighted
with $q^\textup{\# boxes}$ is the simplest model of 
a melting crystal near its corner. We review the 
transfer matrix approach to this model following 
\cite{OR}. This approach can be easily generalized to allow
for certain inhomogeneity and periodicity, which 
is useful in the context of more general models
discussed in Section \ref{sPD}. 

\begin{figure}[!hbtp]
  \begin{center}
    \scalebox{0.4}{\includegraphics{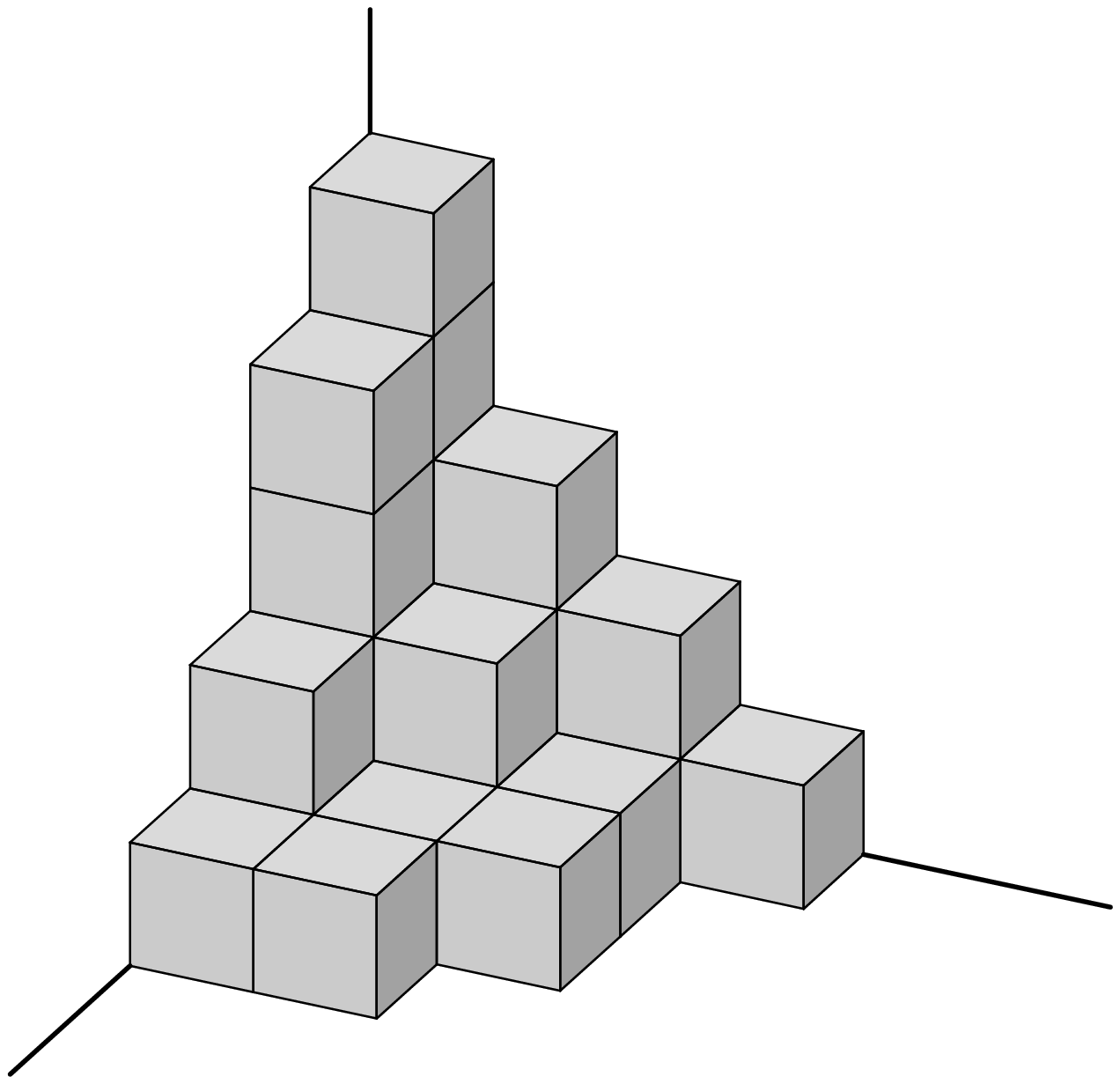}} 
    \scalebox{0.4}{\includegraphics{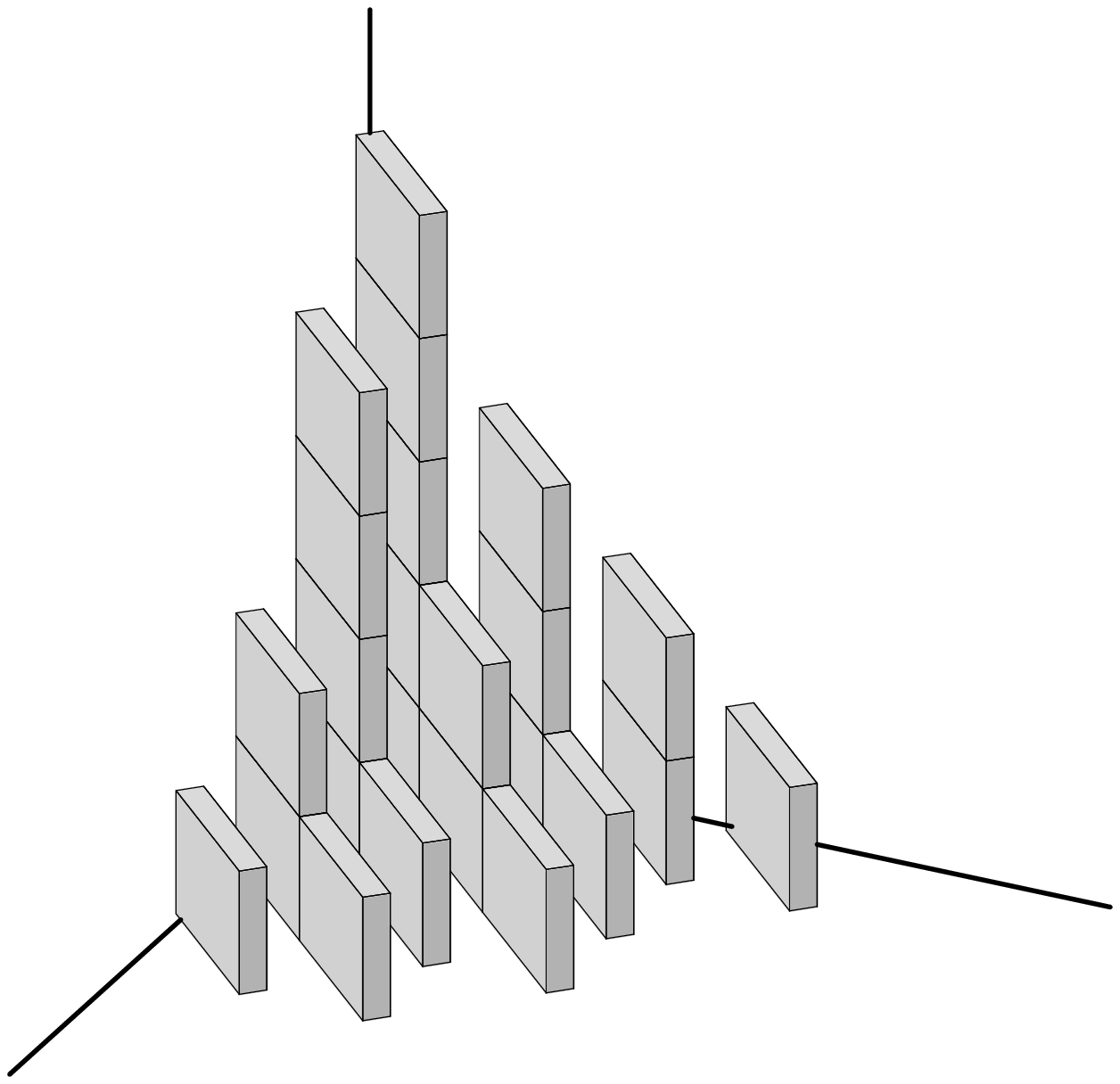}} 
    \caption{A 3d partition and its diagonal slices}
    \label{f3}
  \end{center}
\end{figure}

We start by cutting the 3d partition into diagonal 
slices by planes $x_2- x_1 = t$, see 
Figure \ref{f3}. This operation  makes a 3d partition 
a sequence $\{\mu(t)\}$ of ordinary partitions 
indexed by an integer variable $t$. Conversely,
given a sequence $\{\mu(t)\}$, it can be 
assembled in a 3d partition provided it 
satisfies the following \emph{interlacing}
condition. We say that two partitions $\mu$ and $\nu$
interlace, and write $\mu\succ\nu$ if
$$
\mu_1 \ge \nu_1 \ge \mu_2 \ge \nu_2 \ge \dots \,.
$$
It is easy to see that a sequence of slices 
$\{\mu(t)\}$ of a 3d partition satisfies
$$
\mu(t) \prec \mu(t+1)\,, \quad t<0 \,,
$$
and the reverse relation for $t\ge 0$. 

There is a well-known map from partitions 
to states in the NS sector of 
 complex fermionic oscillator. Let 
$a_i$ and $b_i$ be the areas of pieces 
one gets by slicing a 2d partition first 
diagonally and then horizontally 
(resp.\ vertically) above and the 
below the diagonal, respectively. 
Formally, 
$$
a_i = \mu_i - i + \frac12\,, \quad 
b_i = \mu^t_i -i + \frac12\,,
$$
where $i$ ranges from 1 to the number 
of squares on the diagonal of $\mu$. 
In mathematical literature, these
coordinates on partitions are known
as (modified) Frobenius coordinates. 
The fermionic state associated to $\mu$ is
$$
|\mu \rangle= \prod_{j=1}^d\psi^*_{a_i}\psi_{b_i}
|0\rangle$$
Note that 
$$
q^{L_0}|\mu\rangle =q^{\textup{\# boxes}}|\mu \rangle =
q^{|\mu|}
|\mu \rangle\,,
$$
where we denote the total number of boxes of $\mu$
by $|\mu|$. We can write a 
bosonic representation
of this state by the standard bosonization procedure.  

Consider
the operators
$$
\Gamma_{\pm }(z)=\exp
\left(\sum_{\pm n>0} {z^n J_{n}\over n}\right)\,,
$$
where $J_n$ denotes the modes of the fermionic current
$\psi^* \psi$. The operators $\Gamma_{\pm}(z)$ can be identified
with annihilation and creation parts of the bosonic
vertex operator $e^{\phi (z)}$. The relevance
of these operators for our problem lies in the 
following formulas:
\begin{align} \label{Gpm}
  \Gamma_{-} (1)|\mu\rangle  &=\sum_{\nu{\succ}\mu} |\nu\rangle \\
\Gamma_{+}(1) |\mu \rangle &=\sum_{\nu{\prec}\mu} |\nu \rangle
\notag
\end{align}

To illustrate their power we will now derive, as in
\cite{OR}, the McMahon's  generation function for 
3d partitions
$$
Z = \sum_{\textup{3d partitions $\pi$}} q^\textup{\# of boxes} \,.
$$
By the identification \eqref{Gpm} of the transfer 
matrix, we have 
\begin{equation} \label{Ztr}
  Z = 
\left\langle 
\left(\prod_{t=0}^\infty
q^{L_0}\, \Gamma_+(1)\right) 
q^{L_0}  
\left(\prod_{t=-\infty}^{-1}
 \Gamma_-(1) \, q^{L_0} \right)
\right\rangle \,.
\end{equation}
Now we commute the operators $q^{L_0}$ to the outside, 
splitting the middle one in half. This yields 
\begin{equation}
  Z = 
\left\langle 
\prod_{n>0}
\Gamma_+(q^{n-\frac12})  \, 
\prod_{n>0}
 \Gamma_-(q^{-n-\frac12})
\right \rangle \,.
\end{equation}
Now we commute the creation operators through 
annihilation operators, using the commutation 
relation
$$
\Gamma_+(z)\,\Gamma_-(z')=(1-z/z')^{-1}\,
 \Gamma_-(z')\,\Gamma_+(z)\,.
$$
The product of resulting factors gives directly the 
McMahon's function
$$
Z=\prod_{n>0} (1-q^n)^{-n} =f\,.
$$

Same ideas can be used to get more 
refined results, such as, for example, 
the correlation functions, see \cite{OR}. We
will generalize below the above 
computation to the case when 3d partitions
have certain asymptotic configuration in 
the direction of the three axes. Before doing
this, we will review some relevant 
theory of symmetric functions. 
\subsection{Skew Schur functions}
Skew Schur functions $s_{\lambda/\mu}(x_1,x_2,\dots)$
are certain symmetric polynomials in the 
variables $x_i$ indexed by a pair of 
partitions $\mu$ and $\lambda$ such that 
$\mu\subset\lambda$, see \cite{M}. Their relevance for 
us lies in the well-known fact (see e.g. \cite{Kac})
that 
\begin{equation}
  \label{Gschur}
  \prod_i \Gamma_-(x_i) \, |\mu\rangle = 
\sum_{\lambda \supset \mu} 
s_{\lambda/\mu} (x) \, |\lambda\rangle \,. 
\end{equation}
When $\mu=\emptyset$, this specializes to the 
usual Schur functions. The Jacobi-Trudy
determinantal formula continues to hold for
skew Schur functions:
\begin{equation}
  \label{JT}
  s_{\lambda/\mu} = \det\big( h_{\lambda_i-\mu_j+j-i}\big) \,.
\end{equation}
Here $h_k$ is the complete homogeneous function 
of degree $k$ --- sum of all monomials of degree $k$. 
They can be defined by the generating series
\begin{equation}
  \label{hnt}
  \sum_{n\ge 0} h_n t^n = \prod_i (1- t x_i)^{-1}  =
\exp\left(\sum_{n>0} \frac{t^n}{n} \, \sum_i x_i^n \right) \,.
\end{equation}
The formula \eqref{JT} is very efficient for 
computing the values of skew Schur function, including
their values at the points of the form 
\begin{equation}
  \label{qmu}
 q^{\nu+\rho} = \left(q^{\nu_1-1/2},q^{\nu_2-3/2},q^{\nu_3-5/2},
\dots\right) \,, 
\end{equation}
where $\nu$ is a partition. 
In this case the sum over $i$ in \eqref{hnt} becomes
a geometric series and can be summed explicitly. 

There is a standard involution in the algebra of 
symmetric function which acts by 
$$
s_\lambda \mapsto s_{\lambda^t} \,,
$$
where $\lambda^t$ denotes the transposed diagram. 
It continues to act on skew Schur functions in 
the same manner 
$$
s_{\lambda/\mu} \mapsto s_{\lambda^t/\mu^t} \,.
$$
It is straightforward to check that
\begin{equation}
  \label{invs}
  s_{\lambda/\mu} (q^{\nu+\rho}) = (-1)^{|\lambda|-|\mu|} 
s_{\lambda^t/\mu^t} (q^{-\nu-\rho}) \,.
\end{equation}

Finally, the following property of the skew Schur
function will be crucial in making the connection 
to the formula for the topological vertex from 
\cite{Aga}. The coefficients of the expansion 
\begin{equation}\label{expnd}
  s_{\lambda/\mu}  = \sum_\nu c^{\lambda}_{\mu\, \nu} s_\nu
\end{equation}
of skew Schur functions in terms of ordinary Schur
function are precisely the tensor product 
multiplicities, also known as the Littlewood-Richardson 
coefficients. 
\subsection{Topological vertex and 3d partitions}
\subsubsection{Topological vertex in terms of Schur functions}
Our goal now is to recast the topological
vertex \cite{Aga} in terms of Schur functions.  The basic
ingredient of the topological vertex involves the expectation
values of $U(\infty)$ Chern-Simons Hopf link invariant $W_{
\mu \lambda}=W_{\lambda \mu}$ in representations $\mu$ and $\lambda$.
The expression of the 
 Hopf link invariant $W_{\mu\lambda}$ in terms of the 
Schur functions is the following: 
\begin{equation}
  \label{Wml}
  W_{\mu\lambda} = W_\mu \, s_\lambda(q^{\mu+\rho})\,,
\end{equation}
where
\begin{equation}
  \label{Wm}
  W_\mu = q^{\kappa(\mu)/2} \, s_{\mu^t} (q^\rho) \,. 
\end{equation}
Here
\begin{equation}
  \label{kap}
  \kappa(\lambda) = \sum_i 
\left[\left(\lambda_i-i+\tfrac12\right)^2 - \left(-i+\tfrac12
\right)^2\right] =  2 \sum_{\square=(i,j)\in \lambda} (j-i) 
\end{equation}
is the unique up to scalar quadratic Casimir such that
$$
\kappa(\emptyset) = \kappa(\square) = 0\,.
$$
Using \eqref{expnd} it is straightforward to check that 
the topological vertex $C(\lambda,\mu,\nu)$
in the standard framing has the following expression in 
terms of the skew Schur functions
\begin{equation}
  \label{topver}
  C(\lambda,\mu,\nu) = q^{\kappa(\lambda)/2+\kappa(\nu)/2} 
s_{\nu^t}(q^{\rho}) \, \sum_\eta s_{\lambda^t/\eta} (q^{\nu+\rho}) 
\, s_{\mu/\eta} (q^{\nu^t+\rho}) \,.
\end{equation}
\subsubsection{The lattice length}
To relate the crystal melting problem to the topological
vertex we first have to note that the topological vertex
refers to computations in the A-model corresponding
to placing Lagrangian branes on each leg of ${\bf C}^3$, assembled
into representations of $U(\infty)$ and identified with
partitions.  If we place the brane at a fixed position
and put it in a representation $\mu$, the effect of moving
the brane from a position $l$ to the position $l+k$
 (in string units) affects
the amplitude by a multiplication of
$$\exp(-k|\mu|)$$
Now consider the lattice model where we fix the asymptotics
at a distance $L>>1$ to be fixed to be a fixed partition $\mu$.
Then if we change $L\rightarrow L+K$ then the amplitude
gets weighted by $q^{K|\mu|}$. Since we have
already identified $q=e^{-g_s}$ 
 Comparing these
two expressions we immediately deduce that
$$k=Kg_s$$
In other words the distance in the lattice computation
times $g_s$ is the distance as measured in string units.
This is satisfactory as it suggests that as $g_s\rightarrow 0$
the lattice spacing in string units goes to zero and the space
becomes continuous.  

In defining the topological
vertex one gets rid of the propagator factors above (which
will show up in the gluing rules). Similarly in the lattice
model when we fix the asymptotic boundary condition to be given
by fixed 2d partitions we should multiply the amplitudes
by $q^{-L |\mu|}$ for each fixed asymptote at lattice
position $L$.  Actually  this is not precisely right:
we should rather counter weight it with $q^{-(L+{1\over 2})|\mu|}$.
To see this note that if we glue two topological vertices with
lattice points $L_1$ and $L_2$ along the joining edge, the number of
points along the glued edge is $L_1+L_2+1$.  Putting the ${1\over 2}$
in the above formula gives a symmetric treatment of this issue
in the context of gluing.
\subsubsection{Framing}
Topological vertex also comes equipped with a framing \cite{Aga}\
for each edge, which we now recall. Toric Calabi-Yau's
come with a canonical direction in the toric base.  In the
case of ${\bf C}^3$, it is the diagonal line $x_1=x_2=x_3$ in the
$O^+$.  One typically projects vectors on this toric
base along this direction, to a 2-dimensional plane (the $U,V$
plane in the context we have discussed). At each vertex there
are integral projected 2d vectors along the axes which sum up to zero.
 The topological
vertex framing is equivalent to picking a 2d projected
vector on each axis whose cross product with the integral
vector along the axis is $+1$ (with a suitable sense of orientation).
If $v_i$ is a framing vector for the $i$-th axis, and $e_i$
denotes the integral vector along the $i$-th axis, then the
most general framing is obtained by
$$v_i\rightarrow v_i+n_ie_i$$
where $n_i$ is an integer.  We now interpret this choice
in our statistical mechanical model:  In describing
the asymptotes of the 3d partition we have to
choose a slicing along each axis.  We use the framing vector,
together with the diagonal direction $x_1=x_2=x_3$ 
to define a slicing
2-plane for that edge.  The standard framing corresponds to
choosing the framing vector in cyclic order:  On the $x_1$-axis,
we choose $x_3$, on the $x_3$-axis we choose $x_2$ and on the $x_2$-axis
we choose $x_1$. This together with the diagonal line
determines a slicing plane on each axis.  Note that
all different slicings will have the diagonal line on them.
This line passes through the diagonal of the corresponding
2d partition.

  Before
doing any detailed comparison with the statistical mechanical
model with fixed asymptotes we can check whether framing dependence of the 
topological
vertex can be understood.  This is indeed the case.  Suppose
we compute the partition function with fixed 2d asymptotes and
with a given framing (i.e. slicing).  Suppose we shift the framing
by $n_i$.  This will still cut the asymptotic diagram along the same
2d partition.  Now, however, the total number of boxes of the 3d
partition has changed.  The diagonal points of the partition
have not moved as they are on the slicing plane
for each framing.  The farther a point is from the diagonal
the more it has moved.  Indeed the net number of points added
to the 3d partition is given by
$$n_i\sum_{k,l\in \mu} (k-l)=n_i \kappa(\mu)$$
Thus the statistical mechanical model will have the extra Boltzmann
weight $q^{n_i \kappa (\mu)}$.  This is precisely the framing
dependence of the vertex.  Encouraged by this observation
we now turn to computing the topological vertex in the standard
framing from the crystal point of view.

\subsection{The perpendicular partition function}
\subsubsection{Definition}\label{sdefP}
Now our goal is to find an exact match between 
the formula \eqref{topver}
and the partition function $P(\lambda,\mu,\nu)$ for 3d partitions 
whose asymptotics in the direction of the three
coordinates axes is given by three given partitions
$\lambda$, $\mu$, and $\nu$. 
This generating function, which we call the 
\emph{perpendicular partition function}, will be 
defined and computed presently. 

\begin{figure}[!hbtp]
  \begin{center}
    \scalebox{0.5}{\includegraphics{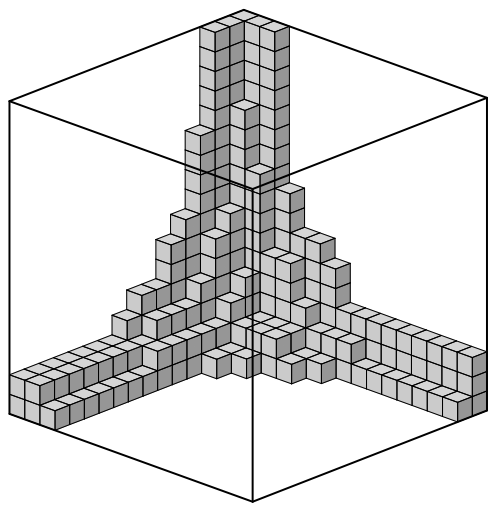}} 
    \caption{A 3d partition ending on three given 2d partitions}
    \label{f4}
  \end{center}
\end{figure}

Consider 3-dimensional partitions $\pi$ inside the box
$$
[0,N_1] \times [0,N_2] \times [0,N_3] \,.
$$
Let the boundary conditions in the planes 
$x_i = N_i$ be given by three partitions 
$\lambda$, $\mu$, and $\nu$. This means 
that, for example,  the facet of $\pi$ 
in the plane $x_1=N_1$ is the diagram of the
partition $\lambda$ oriented so that $\lambda_1$
is its length in the $x_2$ direction. The other
two boundary partitions are defined in the 
cyclically symmetric way. See Figure \ref{f4}
in which $\lambda=(3,2)$, $\mu=(3,1)$ and $\nu=(3,1,1)$. 

Let $P_{N_1,N_2,N_3}(\lambda,\mu,\nu)$ be the
partition function in which every $\pi$ is 
weighted by $q^{\textup{vol}(\pi)}$. It is 
obvious that the limit 
\begin{equation}
  \label{P}
  P(\lambda,\mu,\nu) = 
\lim_{N_1,N_2,N_3\to\infty} q^{-N_1 |\lambda| - N_2 |\mu| - N_3 |\nu|} 
P_{N_1,N_2,N_3}(\lambda,\mu,\nu)
\end{equation}
exists as a formal power series in $q$. What should the relation
of this to topological vertex be?  From what we have said before
this should be the topological vertex itself up to framing factors.
Let us also fix the framing factor, to compare it to the canonical 
framing.  First of all we need to multiply $P$ by 
$$q^{{-1\over 2}(|\lambda|+|\mu|+|\nu|)}$$
This is related to our discussion of the gluing algorithm
(of shifting $N_i\rightarrow N_i+{1\over 2}$) and
splitting the point of gluing between the two vertices.  Secondly
this perpendicular slicing is not the same as canonical framing.
We need to rotate the perpendicular slicing to become
the canonical framing.  This involves the rotation of the partition
along its first column by one unit and the
sense of the rotation is to increase the number of points.
For each representation this gives 
$${1\over 2} \sum_i \lambda_i (\lambda_i-1)={1\over 
2}(\|\lambda\|^2-|\lambda|)$$
 extra boxes which we have to subtract
off and gives the additional weight.  Here
$$
\|\lambda\|^2 = \sum \lambda_i^2 \,.
$$
Combining
with the previous factor we get a net factor of
$$q^{{1\over 2}(||\lambda||^2+||\mu||^2 +||\nu||^2)}$$
Moreover we should normalize as usual by dividing
by the partition function with trivial
asymptotic partition, which is the
McMahon function.  We thus expect
$$\prod_n (1-q^n)^n q^{{1\over 2}(||\lambda||^2+||\mu||^2 +||\nu||^2)}
P(\lambda ,\mu ,\nu )=C(\lambda ,\mu ,\nu).$$
We will see below that this is true up to $g_s\rightarrow -g_s$
and an 
overall factor
that does not affect the gluing properties of the vertex
(as they come in pairs) and it can be viewed
as a gauge choice for the topological vertex.

\subsubsection{Transfer matrix formula}
\begin{figure}[!hbtp]
  \begin{center}
    \scalebox{0.4}{\includegraphics{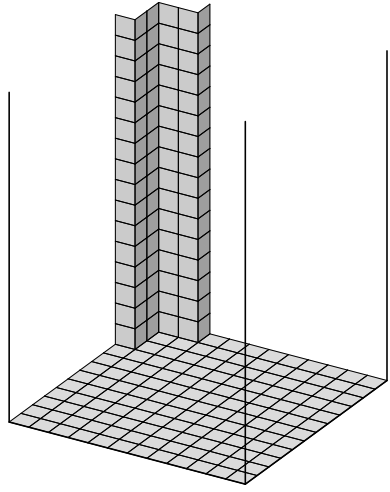}} 
    \caption{The container for skew 3d partitions}
    \label{f8}
  \end{center}
\end{figure}

Recall that in the transfer matrix setup one
slices the partition diagonally. Compared
with the perpendicular cutting, the 
diagonal cutting adds extra boxes to 
the partition and, as a result, it 
increases its volume by $\binom{\lambda}{2}+
 \binom{\mu^t}{2}$, where, by definition
 \begin{equation}
   \label{bilam}
   \binom{\lambda}{2} = \sum_i \binom{\lambda_i}{2} \,.
 \end{equation}
Also, for the transfer-matrix method it is 
convenient to let $N_3=\infty$ from
the very beginning, that is, to consider
\begin{equation}
  \label{P2}
  P_{N_1,N_2}(\lambda,\mu,\nu) = 
\lim_{N_3\to\infty} q^{- N_3 |\nu|} 
P_{N_1,N_2,N_3}(\lambda,\mu,\nu) \,.
\end{equation}
The partition function $P_{N_1,N_2}(\lambda,\mu,\nu)$
counts 3d partitions $\pi$ 
with given boundary conditions on the planes $x_{1,2}= N_{1,2}$
inside the container which is a
 semiinfinite cylinder with base
$$
[0,N_1]\times [0,N_2] \setminus \nu\,,
$$
see Figure \ref{f8}. In other words, $P_{N_1,N_2}(\lambda,\mu,\nu)$
counts \emph{skew 3d partitions} in the sense of \cite{OR}. 

The main observation about skew 3d partitions is that their
diagonal slices interlace in the pattern dictated by the 
shape $\nu$. This gives the following 
 transfer-matrix formula  for \eqref{P2}, which 
is a direct generalization of \eqref{Ztr}: 
\begin{multline}
  \label{opform}
  P_{N_1,N_2}(\lambda,\mu,\nu) =
q^{-\binom{\lambda}{2}-
 \binom{\mu^t}{2}} \, \times \\ 
\left\langle \lambda^t\left|
\left(\prod_{\substack{N_1-1\\ \textup{terms}}}
q^{L_0}\, \Gamma_\pm(1)\right) 
q^{L_0}  
\left(\prod_{\substack{N_2-1\\ \textup{terms}}}
 \Gamma_\pm(1) \, q^{L_0} \right)
\right| \mu \right\rangle \,,
\end{multline}
where the pattern of pluses and minuses in $\Gamma_\pm$ 
is dictated by the shape of $\nu$. 
\subsubsection{Transformation of the operator formula}
Now we apply to the formula \eqref{opform}
the following three transformations:
\begin{itemize}
\item Commute the operators $q^{L_0}$ to the outside,
splitting the middle one in half. The operators
$\Gamma_\pm(1)$ will be conjugated to the operators
$\Gamma_\pm(q^{\dots})$ in the process.
\item Commute the raising operators $\Gamma_-$ to 
the left and the lowering operators $\Gamma_+$ to
the right. There will be some overall, $\nu$-dependent
multiplicative factor $Z(\nu)$ from this operation. 
\item Write the resulting expression as a sum over
intermediate states $|\eta\rangle\langle \eta|$. 
\end{itemize}
\subsubsection{The result}
We obtain
\begin{equation}
  \label{P3}
   P(\lambda,\mu,\nu)= Z(\nu) \, q^{-\binom{\lambda}{2}-
 \binom{\mu^t}{2}-|\lambda|/2-|\mu|/2} \, 
\sum_{\eta} s_{\lambda^t/\eta}(q^{-\nu-\rho}) \, 
s_{\mu/\eta}(q^{-\nu^t-\rho}) \,.
\end{equation}
In order to determine the multiplicative factor $Z(\nu)$, 
we compute
$$
P(\emptyset,\emptyset,\nu) = P(\nu,\emptyset,\emptyset) 
$$
using the formula \eqref{P3}. We get
\begin{equation}
  \label{Znu}
 Z(\nu) = 
\frac{q^{-\binom{\nu}{2}-|\nu|/2} \, s_{\nu^t} (q^{-\rho})}
{\prod_{n>0} (1-q^n)^n} \,. 
\end{equation}
Since 
\begin{equation}
  \label{kappabin}
  \frac{\kappa(\mu)}{2} = \binom{\mu}{2} - \binom{\mu^t}{2}\,,
\end{equation}
comparing \eqref{P3} with \eqref{topver} we obtain
\begin{equation}
  \label{Pver}
  \boxed{  C(\lambda,\mu,\nu;1/q) = q^{\frac{\|\lambda^t\|+\|\mu^t\|+\|\nu^t\|}{2}}
\, \prod_{n>0} (1-q^n)^n \, P(\lambda,\mu,\nu) \quad 
}
\end{equation}
where
\begin{equation}
  \label{nrms}
  \frac{\|\lambda\|^2}{2} = \sum_{i} \frac{\lambda_i^2}{2} = 
 \binom{\lambda}{2}+\frac{|\lambda|}{2} \,. 
\end{equation}
This is exactly what we aimed for, see Section \ref{sdefP},
up to the following immaterial details. The difference
between $\|\lambda^t\|$ and $\|\lambda\|$ is irrelevant
since the gluing formula pairs $\lambda$ with $\lambda^t$. 
Also, since the string amplitudes are even functions of 
$g_s$, they are unaffected by the substitution $q\mapsto 1/q$. 
These differences can be viewed as a gauge choice
for the topological vertex. 

\subsection{Other generalizations}
One can use the topological vertex to glue various
local ${\bf C^3}$ patches and obtain the topological
A-model amplitudes.  Thus it is natural to expect
that there is a natural lattice model.  There is a natural
lattice \cite{inov}\ for the crystal
in this case, obtained by viewing the K\"ahler form divided
by $g_s$ as defining the first Chern class of
a line bundle and identifying the lattice model with holomorphic
sections of this bundle.  This is nothing but geometric
quantization of Calabi-Yau with the K\"ahler form
playing the role of symplectic structure and 
$g_s$ playing the role of $\hbar$. 
The precise definition of lattice melting
problem for this class is currently under investigation \cite{inov}.
The fact that we have already seen the emergence of topological
vertex in the lattice computation corresponding to ${\bf C}^3$ 
one would expect that asymptotic gluings 
suitably defined should give the gluing rules of the statistical
mechanical model.
It is natural to expect this idea also works the same
way in the compact case, by viewing the Calabi-Yau as
a non-commutative manifold with non-commutativity
parameter being $g_sk$, where $k$ is the K\"ahler form.

It is also natural to embed this in superstring (this
is currently under investigation \cite{ov}), where
$g_s$ will be replaced by graviphoton field strength.
The large $g_s$ in this context translates to strong graviphoton
field strength, for which it is natural to expect discretization
of spacetime.  This is exciting as it will potentially
give a
novel realization of superstring target space as a discrete lattice.
\section{Periodic dimers and toric local CY}\label{sPD}
\subsection{Dimers on a periodic planar bipartite graph}
A natural generalization of the ideas discussed here is 
the planar dimer model, see \cite{Ke} for 
an introduction. Let $\Gamma$ be a planar graph 
(``lattice'') which is periodic and bipartite. 
The first condition means that it is lifted from 
a finite graph in the torus $\T^2$ via the 
standard covering map $\R^2\to \T^2$. The 
second condition means that the vertices of 
$\Gamma$ can be partitioned into two 
disjoint subsets (``black'' and ``white'' vertices)
such that edges connect only white vertices
to black vertices. Examples of such graphs
are the standard square or honeycomb lattices. 

By definition, a dimer configuration on $\Gamma$ is a collection 
of edges $D=\{e\}$ such that every vertex is
incident to exactly one edge in $D$. Subject to 
suitable boundary conditions, the partition 
function of the dimer model is defined by
$$
Z= \sum_D \prod_{e\in D} w(e) \,,
$$
where $w(e)$ is a certain (Boltzmann) weight assigned to 
a given edge. For example, one can take both weights
and boundary conditions to be periodic, in which 
case $Z$ is a finite sum. One can also impose
boundary conditions at infinity by saying that the 
dimer configuration should coincide with a given 
configuration outside some ball of a large radius. 
The relative weight of such a configuration is 
a well-defined finite product, but the sum $Z$ 
itself is infinite. In order to make it 
convergent, one introduces a factor of $q^{\textup{volume}(D)}$,
defined in terms of the height function, see 
below. Other boundary conditions can be 
given by cutting a large but finite piece out of the 
graph $\Gamma$ and considering dimers on it. 

Simple dimer models, such as equal weight 
square or honeycomb grid dimer models with 
simple boundary conditions were first 
considered in the physics literature many 
years ago \cite{Kast,TF}. A complete theory of 
the dimer model on a periodic weighted
planar bipartite graph was developed 
in \cite{KOS,KO}. It has some distinctive 
new features due to spectral curve
being a general high genus algebraic 
curve. We will now quote some 
results of \cite{KOS,KO} and indicate their
relevance in our setting. 
\subsection{Periodic configurations and spectral curve}
Consider dimer configuration $D$ on a torus or, 
equivalently, a dimer configuration in the 
plane that repeat periodically, like a 
wall-paper pattern. There are finitely many 
such configurations and they will play a 
special role for us, namely they will 
describe the possible facets of our CY 
crystal. We will now introduce a certain 
refined counting of these configurations. 

Given two configurations $D_1$ and $D_2$ on 
a torus $\T^2$, their union is a collection of 
closed loops on $\T^2$. These loops come with 
a natural orientation by, for example, 
going from white to black vertices along 
the edges of the first dimer and from black
to white vertices along the edges of the second dimer. Hence, they 
define an element (by summing over all classes of the loops)
$$
h= (h_1,h_2) \in H_1(\T^2,\Z) \,,
$$
of the first homology group of the torus. 
Fixing any configuration $D_0$ as our reference
point, we can associate $h=h(D)$ to any 
other dimer configuration and define 
\begin{equation}
  \label{spK}
  F(z,w) = \sum_{D} (-1)^{Q(h)} \, z^{h_1} \, w^{h_2} \, 
\prod_{e\in D} w(e) \,,
\end{equation}
where $Q(h)$ is any of the 4 theta-characteristic, for example,
$Q((h_1,h_2))=h_1 h_2$. The ambiguity in the definition 
of \eqref{spK} comes from the choice of the reference dimer $D_0$,
which means overall multiplication by a monomial in $z$ and $w$,
the choice of the basis for $H_1(\T^2,\Z)$, which means $SL(2,\Z)$
action, and the choice of the theta-characteristic, which 
means flipping the signs of $z$ and $w$. 

The locus $F(z,w)=0$ defines a curve in the toric
surface corresponding to the Newton polygon of $F$. 
It is called the \emph{spectral curve} of the dimer
problem for the given set of weights. It is the 
spectral curve of the Kasteleyn operator on $\Gamma$, 
the variables $z$ and $w$ being the Bloch-Floquet 
multipliers in the two directions. In our situation, 
the curve $F(z,w)$ will be related to the mirror
Calabi-Yau threefold.

\subsection{Height function and ``empty'' configurations}
Now consider dimer configurations in the plane. 
The union of two dimer configurations $D_1$ and $D_2$
again defines a collection of closed oriented loops. 
We can view it as the boundary of the level
sets of a function $h$, defined on the cells 
(a.k.a.\ faces) of the graph $\Gamma$. This 
function $h$ is well-defined up to a constant 
and is known as the \emph{height function}. 
\begin{figure}[!hbtp]
  \begin{center}
    \scalebox{0.4}{\includegraphics{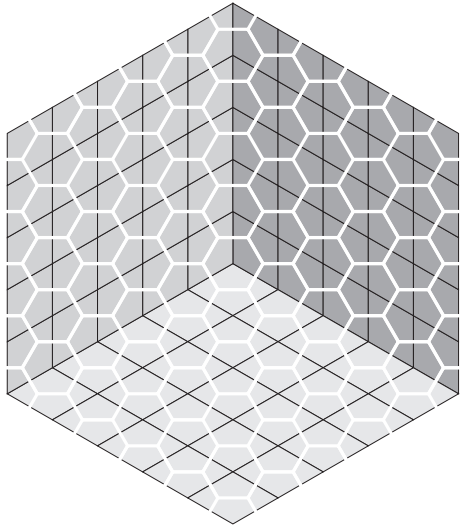}} 
    \caption{The ``Empty room'' configuration of honeycomb dimers}
    \label{f6}
  \end{center}
\end{figure}

It is instructive to see how for dimers on the 
hexagonal lattice this reproduces the 
combinatorics of the 3d partitions, the 
height function giving the previously 
missing 3rd spatial coordinate, see Figure \ref{f6}. The 
``full corner'' or ``empty room'' 
configuration, which was our
starting configuration describing the 
fully quantum $\C^3$ in the language of 
the dimers becomes the unique, up-to 
translation, configuration in which the 
periodic dimer patterns can come together. 
Each rhombus corresponds to one dimer (the edge
of the honeycomb lattice inside it).

\begin{figure}[!hbtp]
  \begin{center}
    \scalebox{0.64}{\includegraphics{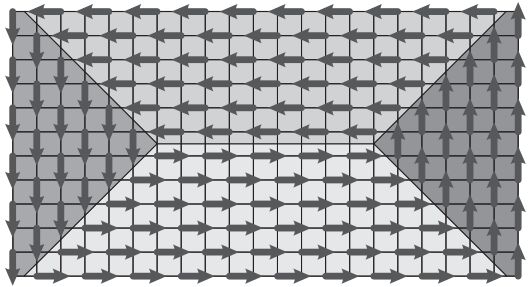}} 
    \caption{An ``Empty room'' configuration of square dimers}
    \label{f7}
  \end{center}
\end{figure}

For general dimers, there are many 
periodic dimer patterns and there
are (integer) moduli in how they
can come together. For example, for the
square lattice, which is the 
case corresponding to the $\cO(-1)\oplus 
\cO(-1)\rightarrow \P^1$ geometry, the periodic patterns
are the brickwall patterns and there 
is one integer degree of freedom in how they 
can be patched together. One possible 
such configuration is shown in Figure \ref{f7}. 
The arrows in Figure \ref{f7} point from white 
vertices to black ones to help visualize the
difference between the 4 periodic patterns. 

 All of these
``empty'' configurations can serve 
as the initial configuration, describing the 
fully quantum toric threefold, for  the 
dimer problem. When 
lifted in 3d via the height function,
each empty configuration follows a 
piece-wise linear function, which is 
the boundary of the polyhedron defining
the toric variety. In particular, the 
number of ``empty room'' moduli 
matches the K\"ahler moduli of the 
toric threefold and changes in its 
combinatorics correspond to the flops. 

Readers familiar with $(p,q)$ 5-brane webs
\cite{ofer}\ and their relation to toric
Calabi-Yau \cite{leung}\ immediately see a dictionary:
A 5-brane configuration is identified as the transition
line from one rigid dimer configuration to other, which
in toric language is related to which $\T^2$ cycle of
Calabi-Yau shrinks over it.  For example the
 the $\C^3$
geometry is mapped to
three 5-branes (of types
$(1,0),(0,1),(-1,-1)$) on the 2 plane meeting at a point where
each region gets identified with a particular dimer configuration.
Similar description holds for arbitrary 5-brane webs.  In this context
the $F(z,w)=F(e^{-u},e^{-v})$ is identified with the mirror geometry
\cite{hv}.\footnote{More precisely as in \cite{hv,hiqv}\
this corresponds to a LG theory with $W=e^{-Y_3}(F(u,v))$ or to
a non-compact CY given as a hypersurface: $\alpha \beta -F(u,v)=0$.}
\subsection{Excitations and limit shape}
Now we can start ``removing atoms'' from the ``full 
crystal'' configuration, adding the cost of $q$ to 
each increase in the height function. The limit
shape that develops, is controlled by the 
\emph{surface tension} of the dimer problem. 
This is a function of a slope measuring
how much the dimer likes to have height
function with this slope. Formally, it is 
defined as the $n\to\infty$ limit of the 
free energy per fundamental domain 
for dimer configurations on the $n\times n$ torus
$\T^2$ restricted to lie in a given homology 
class or, equivalently, restricted to have certain 
slope when lifted to a periodic configuration
in the plane.
 
One of the main results of \cite{KOS} is the 
identification of this function with the 
Legendre dual of the \emph{Ronkin function} 
of the polynomial \eqref{spK} defined by
\begin{equation}
  \label{RnkF}
  R(U,V) = \frac1{4\pi^2} \iint_{0}^{2\pi} 
\log \, |F(U+ i \theta, V+i \phi)|\, d\theta d\phi\,.
\end{equation}
The Wulff construction implies that 
the Ronkin function itself is
one of the possible limit shapes, the one
corresponding to its own boundary conditions.
Note that this is exactly what one would anticipate
from our general conjecture if we view $F(z,w)$ as the
describing the mirror geometry. In fact following
the same type of argument as in the ${\bf C}^3$ case 
discussed before, would
lead us to the above Ronkin function. 

\begin{figure}[!hbtp]
  \begin{center}
    \scalebox{0.64}{\includegraphics{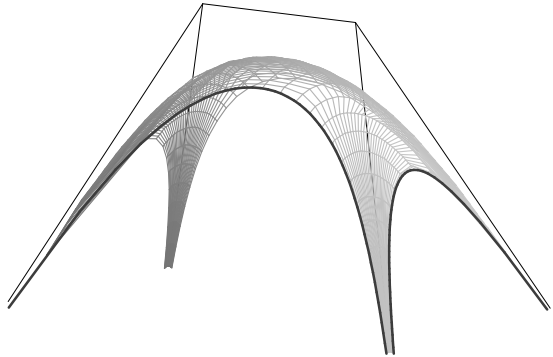}} 
    \caption{The Ronkin function of a hyperbola}
    \label{f5}
  \end{center}
\end{figure}

As an example consider the dimers on the hexagonal 
lattice with $1\times 1$ fundamental domain.  In this case
the edge weights can be gauged away and 
\eqref{RnkF} becomes the Ronkin function 
considered in Section \ref{MCY}.  For the 
square lattice with $1\times 1$ fundamental 
domain, there is 1 gauge invariant 
combination of the 4 weights and the 
spectral curve takes the form
$$
F(z,w) = 1 + z + w - e^{-t} zw \,,
$$
where $t$ is a parameter related to 
the size of $\P^1$ in the $\cO(-1)\oplus 
\cO(-1)\rightarrow \P^1$ geometry. The spectral curve
is a hyperbola in $\C^2$ and (the negative of)  its
Ronkin function is plotted in Figure \ref{f5}. 
Note how one can actually see the projection of the mirror curve ! 

All possible limit shapes for given set of
dimer weights are 
maximizers of the surface tension functional 
and in this sense are very similar to minimal 
surfaces. An analog of the Weierstra\ss\
parameterization for them in terms
of analytic data was found in \cite{KO}. 
It reduces the solution of the Euler-Lagrange
PDE's to solving equations for finitely
many parameters, essentially
finding a plane curve of given degree and 
genus satisfying certain 
tangency and periods conditions. 

In the limit of extreme weights and large 
initial configurations, the amoebas and 
Ronkin functions degenerate to the 
piecewise-linear toric geometry. In this 
limit, it is possible to adjust 
parameters to reproduce the
topological vertex formula for the 
GW invariants of the toric target
obtained in \cite{Aga}. We expect that the general case
will reproduce the features of the 
background dependence (i.e. holomorphic
anomaly) in the A-model
\cite{bcov}. This issue is presently under
investigation \cite{kov}. 

\newpage 

\bigskip 
\noindent{\large \textbf{Acknowledgments}}
\medskip 

We would like to thank the 
Simons workshop in Mathematics and Physics at Stony Brook for
a very lively atmosphere which directly led to this work.
We would also like to thank M.~Aganagic, R.~Dijkgraaf, S.~Gukov,
A.~Iqbal, S.~Katz, A.~Klemm, M.~Marino, N.~Nekrasov, H.~Ooguri, M.~Rocek, G.~Sterman
and A.~Strominger  for valuable discussions. We thank 
J.~Zhou for pointing out an inaccuracy in the formula
\eqref{topver} in the first version of this paper.

A.~O.\ was partially supported by 
DMS-0096246 and fellowships from the Packard foundation. N.~R.\
was partially supported by the NSF grant DMS-0307599.
The research of C.V. is supported in part by NSF grants
PHY-9802709 and DMS-0074329.

\end{document}